\documentclass{webofc}
\usepackage[varg]{txfonts}   % Web of Conferences font
\usepackage{hyperref}
\usepackage{amsmath}
\usepackage{graphicx}
\usepackage{caption}
\usepackage{url}
\usepackage{xspace}
\usepackage{comment}
\excludecomment{confidential}

\hypersetup{colorlinks=true,citecolor=blue,urlcolor=blue,linkcolor=blue}
\begin{document}
\title{New insights into strange-quark hadronization measuring multiple (multi-)strange hadron production in small collision systems with ALICE}
%
% subtitle is optional
%
%%%\subtitle{Do you have a subtitle?\\ If so, write it here}

\author{\firstname{\textit{Sara}} \lastname{Pucillo}\inst{1}\fnsep\thanks{\email{sara.pucillo@cern.ch}} for the ALICE Collaboration }

\institute{Dipartimento di Fisica, Università degli Studi di Torino, and Sezione INFN Torino}

\abstract{%Among the most important results from Run-1 and Run-2 at the LHC is the observation of enhanced production of (multi-)strange to non-strange hadron yields, gradually rising from low-multiplicity to high-multiplicity pp and p--Pb collisions, reaching values close to those measured in peripheral Pb--Pb collisions. More insightful information about the production mechanism could be provided by measuring the (multi-)strange particle multiplicity distribution, P($\textit{n}_{S}$), using a novel method based on counting the number of strange particles event by event. This measurement extends the study of strangeness production beyond its average and represents a new test bench for production mechanisms, probing events with a large imbalance between strange and non-strange content. In this contribution, ALICE results on${\rm K}_{\rm S}^{0}$, $\Lambda$, $\Xi^{-}$ and $\Omega^{-}$ (together with antiparticle) multiplicity distributions in pp collisions at $\sqrt{s}$ = 5.02 TeV as a function of the charged particle multiplicity, together with the average probability for the production multiplets are presented. The results are compared to state-of-the-art phenomenological models implemented in commonly used Monte Carlo event generators, drastically enhancing the sensitivity to the different processes implemented in each approach.
Among the most important results from the Run-1 and Run-2 of the LHC is the observation of an enhanced production of (multi-)strange to non-strange hadron yields, gradually rising from low-multiplicity to high-multiplicity pp and p--Pb collisions, reaching values close to those measured in peripheral Pb--Pb collisions \cite{nature}. The observed behaviour cannot be quantitatively reproduced by any of the available QCD-inspired MC generators. In this contribution an extension of this study is presented: the measurement of the ${\rm K}_{\rm S}^{0}$, $\Lambda$, $\Xi^{-}$ and $\Omega^{-}$ (together with their antiparticles) multiplicity distributions in pp collisions at $\sqrt{s}$ = 5.02 TeV as a function of the charged-particle multiplicity, together with the average probability for the multiplets production, extending the study of strangeness production beyond its average. This novel method, based on counting the number of strange particles event-by-event, represents a new test bench for production mechanisms, probing events with a large imbalance between strange and non-strange content.}
\maketitle
\section{Introduction} \label{intro}
Among several probes of the quark--gluon plasma (QGP) formation, the so-called \textit{Strangeness Enhancement} (SE) was one of the first proposed \cite{rafelski} and observed experimentally \cite{na57}. In addition to confirm the SE at the highest center-of-mass energy \cite{PbPb2.76,pPbV0,pPbCasc}, the ALICE Collaboration carried out a comprehensive study of strange hadron production (relative to charged pions) as a function of the charged-particle multiplicity at midrapidity \cite{nature}. The main finding is that strangeness increases progressively with multiplicity in a compatible way across different collision systems and center-of-mass energies \cite{nature,pp7,pp13}. The invariance of the SE pattern on the colliding system suggests that the mechanisms at play in high-multiplicity pp interactions could be the same as those involved in particle formation in Pb--Pb collisions. Moreover the enhancement, previously unexpected in pp interactions, is proportional to the strangeness content in the hadron being the highest for the $\Omega^{-}$ baryon. However, the observed behaviour cannot be quantitatively reproduced by any of the available QCD-inspired models, suggesting that further developments are needed to obtain a complete microscopic understanding of strangeness production.

\section{Measurement of (multi-)strange particle multiplicity distribution} \label{Pns}
\begin{figure}
    \centering
    \includegraphics[width=0.45 \textwidth, height=0.45\linewidth]{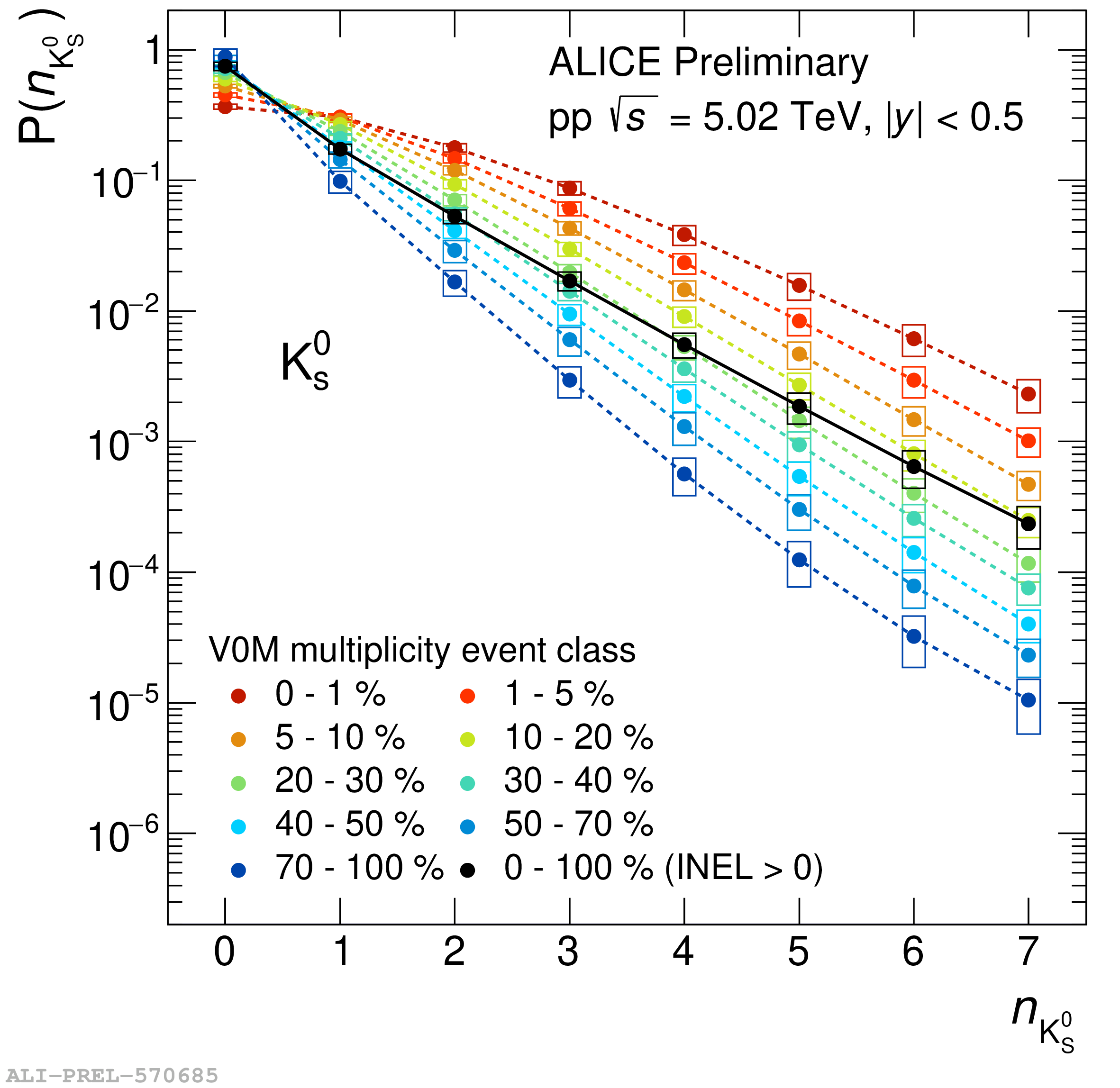} 
    \hspace*{0.3cm}
    \includegraphics[width=0.45 \textwidth, height=0.45\linewidth]{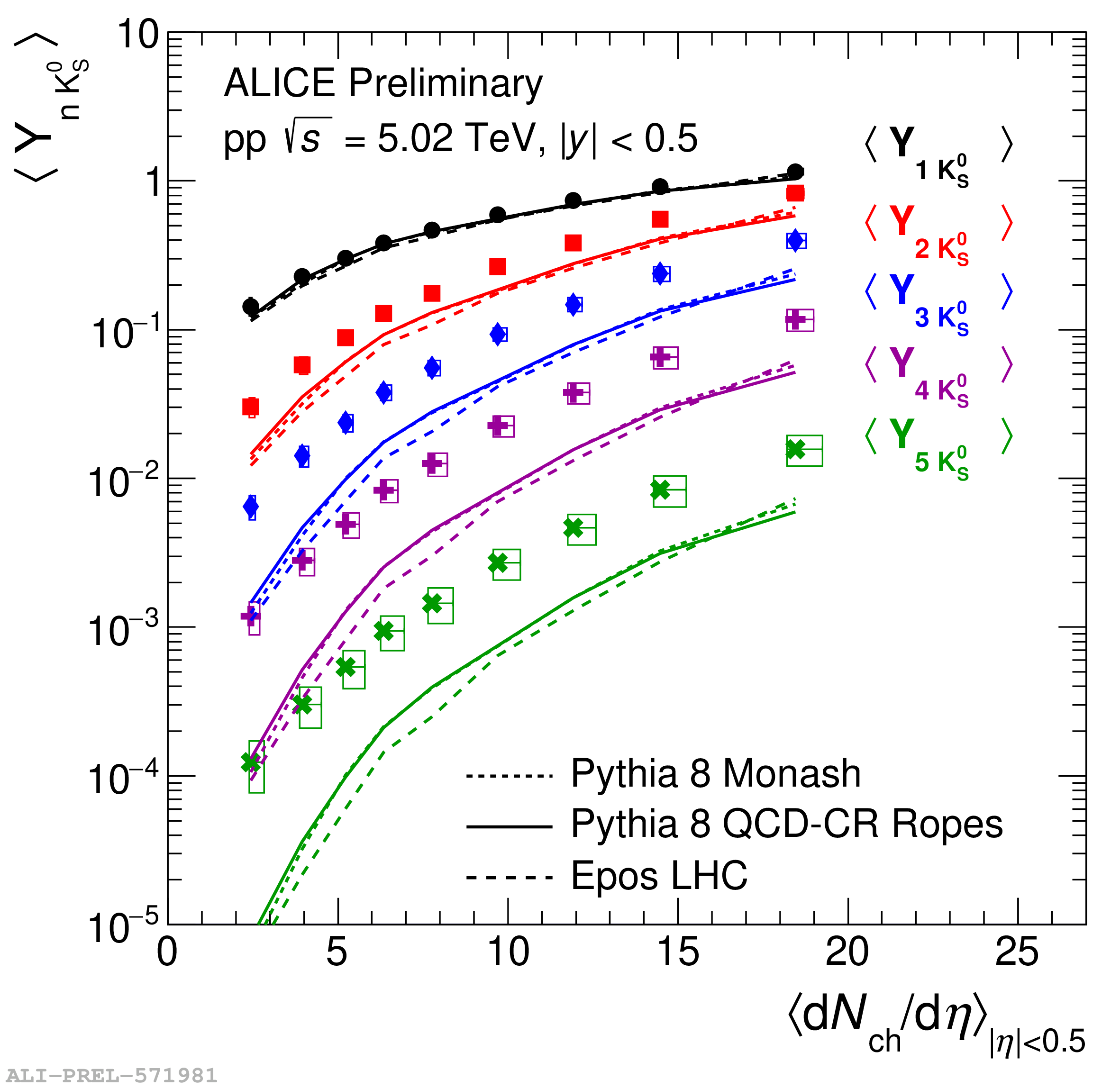}
    \caption{$\textit{Left}$: ${\rm K}_{\rm S}^{0}$ particle multiplicity distribution for several multiplicity classes. The distribution obtained for INEL $>$ 0 has been reported in black and lines are drawn to guide the eye. $\textit{Right}$: Multiple strange hadron production yields for ${\rm K}_{\rm S}^{0}$ a function of the charged-particle multiplicity with model comparison. Model predictions obtained considering Pythia 8 Monash, Pythia 8 QCD-CR Ropes and Epos LHC are shown as dotted, continuous and dashed lines respectively.}
    \label{fig:Distributions}
\end{figure}

To obtain more information about the strangeness production mechanism, the (multi-)strange particle multiplicity distribution, P($\textit{n}_{S}$), has been measured in pp collisions at $\sqrt{s}$ = 5.02 TeV using a new technique based on counting the number of strange particles on an event-by-event basis. \newline
The signal extraction has been performed by developing a procedure based on signal/background weights obtained from the data after the application of selective cuts on topological features of the weak decay and performing particle identification of the daughters. As a first step, a fit to the 1-dimensional invariant mass spectrum given by the sum of a function for the signal (double-sided Crystal Ball \cite{crystallball}) and one for the background (polynomial of order 1 for all particles, except for $\Omega^{-}$ for which a polynomial of order 2 was chosen) was performed in different {\ensuremath{p_{\rm T}}\xspace} and multiplicity bins. In this way, for each invariant mass and {\ensuremath{p_{\rm T}}\xspace} range, it is possible to define a probability that the candidate is signal as the ratio between the value of the signal function and the total one, or background as the ratio between the value of the background function and the total one. After this first step, all events are re-spell-checked to apply the procedure to all the N candidates per event. Event by event, each candidate is considered, associating the probability for it to be signal or background (from the previous step). Then, all the products of the different probabilities associated to each candidate are computed and grouped according to the total signal yield. Consequently, it is possible to obtain that all candidates are signal, all are background, or all the intermediate situations, having for each event with N candidates a full raw probability spectrum spanning from 0 to N. Summing the result obtained for each event, it is possible to measure the spectra for each multiplicity class. In the $\Lambda$ and $\overline{\Lambda}$ analysis, a feed-down correction is performed removing from the total candidates yield the fraction of $\Lambda$ ($\overline{\Lambda}$) coming from the decay of $\Xi^{-}$ ($\overline{\Xi}^{+}$) and $\Xi^{0}$. %The fraction is estimated for each multiplicity class from MC and then is corrected for the {\ensuremath{p_{\rm T}}\xspace} distribution extracted from data, so a weight given by the ratio between the {\ensuremath{p_{\rm T}}\xspace} distribution extracted from data for a specific multiplicity class and the {\ensuremath{p_{\rm T}}\xspace} distribution obtained from MC for the integrated multiplicity is applied.  
The correction for the detector response was obtained by performing a uni-dimensional Bayesian unfolding procedure \cite{unf} using a MC simulation that includes realistic strange particle {\ensuremath{p_{\rm T}}\xspace} distributions and detector conditions.\newline
The probability of producing \textit{n} ${\rm K}_{\rm S}^{0}$ event is reported on the left in Fig. \ref{fig:Distributions} for several multiplicity classes (including INEL $>$ 0 \footnote{INEL $>$ 0 is an event class in which at least one reconstructed silicon pixel detectors tracklet is produced in the pseudorapidity interval $|\eta| <$ 1 with respect to the beam, corresponding to about 75$\%$ of the total inelastic cross-section. }) showcasing that events with up to 7 ${\rm K}_{\rm S}^{0}$ can be observed. The same procedure has been applied to obtain the (multi-)strange particle (antiparticle) multiplicity distribution for $\Lambda$ ($\overline{\Lambda}$), $\Xi^{-}$ ($\overline{\Xi}^{+}$) and $\Omega^{-}$ ($\overline{\Omega}^{+}$) measuring up to 5, 4 and 2 particles/event respectively and observing that, as expected at the LHC energies, the production of particles and antiparticles is the same from the highest to the lowest multiplicity class. It is worth noting that the measurement of the probability of producing \textit{n} particles of a given species per event represents a unique opportunity to test the connection between charged and strange particle multiplicity production all the way to very extreme situations, e.g. spanning from events with 7 ${\rm K}_{\rm S}^{0}$ at low average charged-particle multiplicity -- where $\big<$d\textit{N}/d$\eta \big>_{|\eta|<0.5} \sim$ 3, with potential large fluctuations -- to events with 0 ${\rm K}_{\rm S}^{0}$ high multiplicity -- where $\big<$d\textit{N}/d$\eta \big>_{|\eta|<0.5} \sim$ 20.
%\ensuremath{\langle d\texit{N}/d\eta \rangle_{|\eta|<0.5} \sim 3}, with potential large fluctuations -- to events with 0 ${\rm K}_{\rm S}^{0}$ high multiplicity -- where \ensuremath{\langle d\texit{N}/d\eta \rangle_{|\eta|<0.5} \sim 20}.

\section{Multiple strange hadron production yields and yield ratios} \label{yields}
From the measurement of P($\textit{n}_{S}$) and using Eq. \ref{eq:yieldseq}, it is possible to calculate the average probability $\big< Y_{n {\rm -part}} \big>$ for the production of $n$ particles per event. %that contains the sum of a combinatorial factor and the probability to produce $i-$particles per event of a given species. 
On the right side of Fig. \ref{fig:Distributions}, the average production yield of 1, 2, 3, 4 and 5 ${\rm K}_{\rm S}^{0}$ a function of the charged-particle multiplicity at midrapidity are reported respectively as black, red, blue, magenta and green markers. The increase with multiplicity is more than linear for the production of multiple strange hadrons and comparing the results with Pythia 8 Monash \cite{Pythia}, Pythia 8 (QCD-CR) Ropes \cite{PythiaRopes} and Epos LHC \cite{EPOS}, one can see that the agreement worsens as the number of particles per event increases.

\begin{equation}
  \big< Y_{n {\rm -part}} \big> = \sum_{i=0}^{\infty} \frac{i!}{n!(i-n)!}P_{i-part}.
  \label{eq:yieldseq}
\end{equation}

\begin{figure}
    \begin{center}
    \includegraphics[width=0.95 \textwidth]{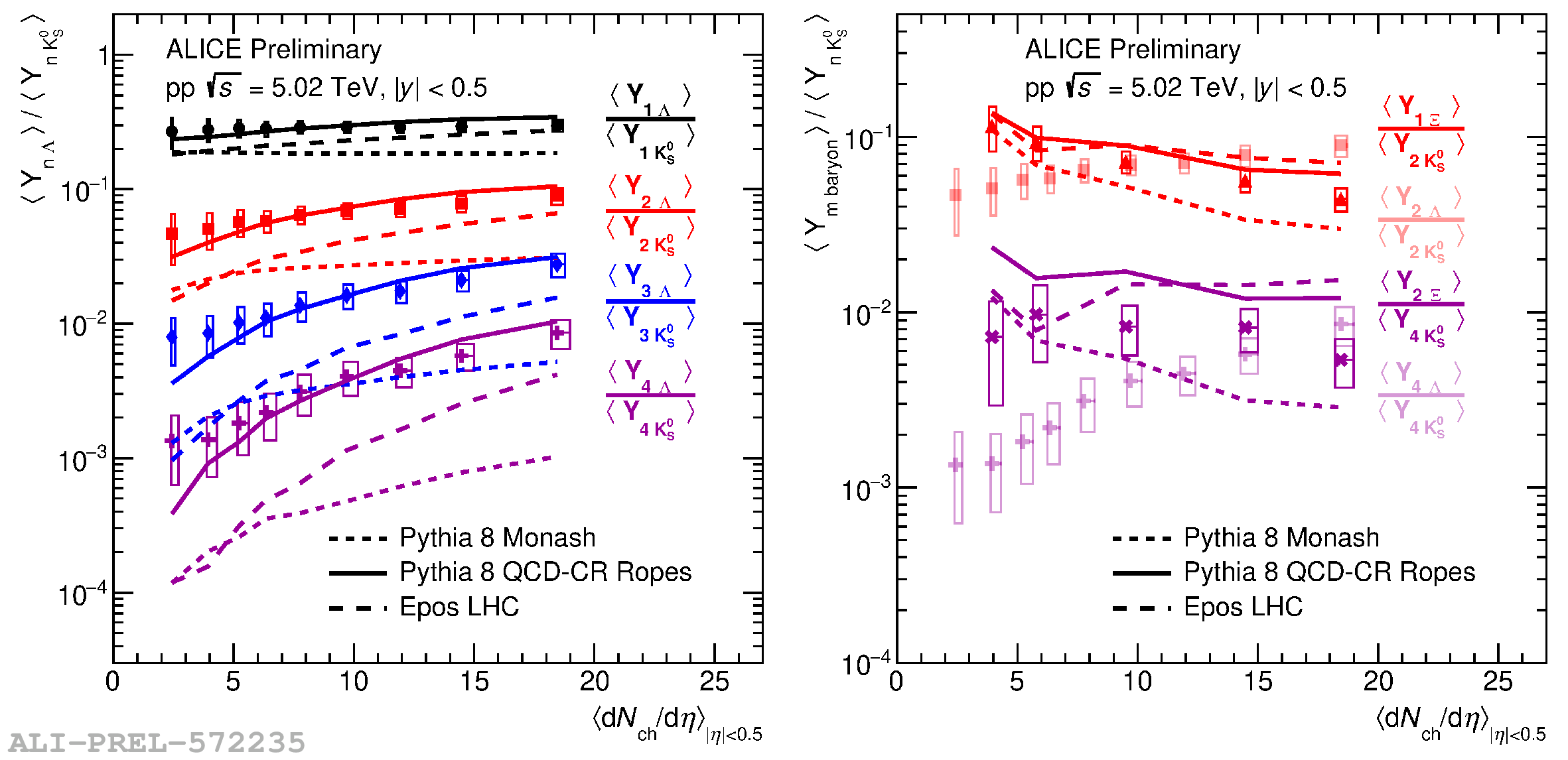}
    \end{center}
    \caption{\textit{Left.} Ratio $\big< Y_{n {\rm \Lambda}} \big> / \big< Y_{n {{\rm K}_{\rm S}^{0}}} \big>$ as a function of the charged-particle multiplicity. \textit{Right.} Ratio $\big< Y_{m {\rm baryons}} \big> / \big< Y_{n {{\rm K}_{\rm S}^{0}}} \big>$ as a function of the charged-particle multiplicity. In both plots there is a comparison to models: Pythia 8 Monash (dotted line), Pythia 8 QCD-CR Ropes (continuous line) and Epos LHC (dashed line).}
    \label{fig:ratiosS0}
\end{figure}

\noindent Thanks to the measurement of multiple strange hadron production yields, one can explore several yield ratios to investigate the relative probability for a specific number of \textit{s}-quarks to hadronize into different types of final hadrons. Firstly, yield ratios with a perfect strangeness balance ($\Delta$S = 0) as a function of the charged-particle multiplicity were considered, starting from the ratio between \textit{n} $\Lambda$ and \textit{n} ${\rm K}_{\rm S}^{0}$. This is reported in the left side of Fig. \ref{fig:ratiosS0}, where an increasing pattern is observed when looking at multiple strange particle production, indicating that when the strangeness remains constant, the likelihood to obtain a baryon and not a meson rises with multiplicity. %These observations suggest that in every strange-hadron/$\pi$ \cite{nature, pp13} plot as a function of multiplicity the observed enhancement can be partly attributed to strangeness and partly to baryon number. 
However, to better understand whether the observed increase is related to the production of the baryon itself or to the difference in mass, type or number of light quarks involved, other multiple strange yield ratios with perfect strangeness balance were considered. Figure \ref{fig:ratiosS0} (right) shows the yield ratios between \textit{m} baryons and \textit{n} ${\rm K}_{\rm S}^{0}$, where a decreasing trend is observed when a larger number of light quarks are involved in the denominator. This is consistent with the idea that at high multiplicity it is simpler to pair \textit{s} with \textit{u} and \textit{d} quarks, which are present in large abundance. Contrarily, in low multiplicity situations the surplus of \textit{s}-quarks (fixed at 2 or 4 for the reported ratios) increases the probability of forming a multi-strange baryon. Looking at the model comparison, both Pythia 8 QCD-CR Ropes and Epos LHC can reproduce qualitatively the increasing or decreasing trends with multiplicity, but Epos LHC tends to underestimate the magnitude. Looking more closely, all trends are rather well reproduced by Pythia 8 QCD-CR Ropes suggesting that the strange quark production rate remains a puzzle, but once \textit{s} is created the model of re-connection with light quarks catches the trends observed in the data. \newline

\noindent Different yield ratios between \textit{m} multi-strange baryons and \textit{n} ${\rm K}_{\rm S}^{0}$, investigating for the first time a difference in the strangeness content between the numerator and the denominator greater than three (up to five), are under preparation and will allow to study the strangeness production at its extremes.

% BibTeX or Biber users please use (the style is already called in the class, ensure that the "woc.bst" style is in your local directory)
% \bibliography{your_bib_file} % Replace "your_bib_file" with the actual name of your .bib file
%
% Non-BibTeX users please use
%

\end{document}